# Resonance frequency dependence on out-of-plane forces for square silicon membranes: applications to a MEMS gradiometer

**I. Lucas[1*], R.P. del Real[1], M.D. Michelena[1], V. de Manuel[1], M. Duch[2], J. Esteve[2,] and J.A. Plaza[2]**

[1]*Instituto Nacional de Técnica Aeroespacial (INTA). Carretera Torrejón- Ajalvir, km 4, 28850 Torrejón de Ardoz, Spain.*

[2]*Instituto de Microelectrónica de Barcelona IMB-CNM (CSIC). Campus UAB s/n, 08193 Bellaterra (Barcelona), Spain.*

*Correspondig author

**Dr. Irene Lucas del Pozo**

Leibniz Institute for Solid State and Materials Research Dresden
Institute for Metallic Materials
Helmholtzstraße 20, D-01069

Dresden, Gemany
Tel.: +49 351 4659 748
FAX: +49 351 4659 541
**e-mail: i.lucas.del.pozo@ifw-dresden.de**

**ABSTRACT**

The dynamic properties of membranes have been object of many researches since they can be used as sensor heads in different devices. Some methods have been proposed to solve the problem of determining the resonance frequencies and their dependence on the stress caused by forces applied on the membrane surface. The problem of the vibrating rectangular membrane under a stress caused by a uniform in-plane force is well known. However, the resonance frequency behaviour when the force is out-of-plane instead of in-plane, is not so well understood and documented. A gradiometer which uses a silicon square membrane with a magnet fixed on it as a sensor head has been developed in a previous work. This device reports a quadratic dependence of the frequency on the out-of-plane magnetic force. In this work, simulations to obtain the dependence of the frequency of the fundamental flexural mode on the stress have been performed. It has been studied the influence of in-plane and out-of-plane forces applied to the membrane. As expected, a square root dependence has been found for in-plane forces. Nevertheless, the problem is more complex when out-of plane forces are considered. Out-of-plane forces gives rise to an initial quadratic dependence which turns into a square root dependence from a certain stress value. The quadratic range increases and the rate of change of the frequency decreases as the surface of the magnet fixed on the membrane increases. The study has addressed these problems and both, experimental and simulated results have been compared and a good agreement between experimental and simulated results has been found.

**INTRODUCTION**

The dynamic properties of membranes have been extensively studied since they can be used as a component in different sensors and devices such as pressure microsensors, where membranes are combined with actuators based on PZT (Pb(Zr,Ti)O$_3$) thin films piezoelectrics in the perovskite phase [1], stress magnetoelastic sensors with applications for remote environmental monitoring [2,3] and, as it has been demonstrated recently, even magnetic force sensors or gradiometers [4]. The dependence of its resonance frequencies on the stress has been extensively studied and simulated [5,6] and several methods have been proposed to solve the problem of determining the natural frequencies and mode shapes of membranes [7]. If the problem is focused on the rectangular membrane vibration, the bibliography reports a square root dependence of the resonance frequency $\omega_{nm}$ on the in-plane stress [8,9]:

$$\omega_{nm} = \sqrt{\frac{T}{\rho}\left[\left(\frac{n\pi}{a}\right)^2 + \left(\frac{m\pi}{b}\right)^2\right]} \tag{1}$$

where $T$ is the stress, $\rho$ is the density of the material of the membrane, $a$ and $b$ are the lateral membrane dimensions and $m$ and $n$ are natural numbers. It is important to remark that this dependence, which for values of small enough stress, tends to a linear dependence, is valid for a stress caused by a uniform in-plane force per unit area applied along the membrane four sides. Both, the square root and the linear dependence have been confirmed experimentally [1-3]. However, the dependence of the frequency on an applied out-of-plane force still remains unreported. Since a great number of devices bases its detection on the vibration of a silicon membrane under the effect of different applied forces, this dependence has become important to be studied, reported and well understood. The membrane vibration turns complicated when the thickness is reduced, the geometry modified and/or the stresses are not applied in the in-plane direction, giving different and complex dependences [10]. In spite of its difficulty, some works have treated the problem of membranes with arbitrary interfaces, studying the change of the resonance frequency of these systems [11], although the evolution of this change depending on the stress or the applied force is not considered.

In a recent work, the experimental dependence of the resonance frequency on the stress caused by an out-of-plane force has been reported [4]. In that device, a silicon membrane with a fixed magnet centred on its surface vibrates as a sensor head of a gradiometer. An alternating magnetic field gradient is generated in order to get the resonance of the sensor head. As a result, in this device the resonance frequency of the sensor head changes showing a quadratic dependence when a DC external magnetic field gradient is applied.

In this device, as in the rest of the devices which base its detection in a membrane vibration, the understanding of the frequency dependence on the stress caused by the applied force on the membrane surface, is fundamental in order to control the device response. In this report, the resonance frequency behaviour when the membrane is stressed due to a force applied not in the in-plane direction but in the out-of-plane one has been investigated. Numerical simulations by the Finite Element Method (FEM) have been carried out applying a force $F$ on the surface of a solid cylinder which models the magnet used in the gradiometer. Different simulations have been performed using different membrane geometries. Depending on the surface of the solid cylinder used to perform the simulation, the area where the force is applied will be changed.

Besides, the theoretical results obtained by ANSYS® simulations have been compared with the experimental measurements of the resonance frequency depending on the magnetic force applied to the device.

**EXPERIMENTAL**

The vibration of 5 mm x 5 mm square silicon (100) membranes with 16 μm thickness, both bare and with a solid cylinder fixed at the centre on its surface, has been studied. The solid cylinder fixed on the membrane consists of a SmCo magnet and it has been simulated with different diameter values using the *Finite Element Method* carried out by ANSYS®. The density, Young modulus and Poisson coefficient values used for the silicon membrane and the SmCo solid cylinder are shown in table 1. The ANSYS elements used for the boundary conditions of the membrane and the solid cylinder are 2D (SHELL93) and 3D (SOLID95) respectively. Both element types allow anisotropic elastic properties. 2D instead of 3D elements have been chosen to model the membrane due to its high aspect ratio since 3D elements would introduce a large amount of elements.

The set-up of the gradiometer used to obtain the experimental measurements is depicted in Figure 1. The sensor consists of a cylindrical permanent SmCo magnet attached to a square Si (100) membrane with a tiny drop of epoxy. After pressing the magnet to the membrane surface, the epoxy will cover the whole bottom area of the magnet. The chip shows two parts: the frame (12 mm x 12 mm and 525 μm thick), and the membrane (5 mm x 5 mm and 16 μm thick). The frame is thicker in order to fix the membrane to a holder without breaking it. The commercial magnets used are SmCo of 1.5 mm height and 3 mm of diameter and they show a strong out-of-plane remanent magnetisation of 0.8 T, what means a magnetic moment in this direction of $6.75 \ 10^{-3} \ Am^2$. The magnet fits, centred on the membrane, without touching the frame. Four coils create an alternating magnetic force on the magnet that excites the sensor head at the resonance frequency. When an external magnetic field gradient is applied or a magnetic specimen is approached to the device, the membrane is stressed and therefore the resonance frequency of the system changes (see Figure 2). The variation of the magnet position is then measured by an optoelectronic method. A coil is considered for calibration of the sensor. It is important to remark that the size of the magnet has been chosen taking into account the fact that the larger the magnetization the higher the magnetic alternating force which forces the membrane to vibrate and, in principle, the lower the magnetic moment that can be easily detected.

Figure 3 shows the simulated static (left) and fundamental flexural mode (right) of a square membrane (top) and a square membrane with a centred fixed cylinder (bottom). The simulations consist in a stressed membrane vibrating so that the resonance frequency can be calculated. Considering the actual device (see Figure 1) where the membrane is vibrating, the membrane is simulated using the clamped boundary conditions. The mechanical properties in the direction of the three axis of the membrane are the mechanical properties of the Si [110] and [001] crystallographic directions for the axis of the plane XY and the z-axis respectively. Since in the direction of the edges of the membrane the mechanical properties are the same, no anisotropy effect must be taken into account in this plane. The bare membrane simulations have been carried out using an in-plane force while those of the membrane with a fixed cylinder were simulated under the effect of an out-of-plane force. From these simulations the resonance frequency dependence on the applied force has been obtained. Only the frequency of the fundamental flexural mode has been considered. Three different models, as shown in table 2, have been used to simulate the vibration of the membrane. One of the geometries consists in a punctual force applied just in the centre of the membrane while vibrating. The two remaining geometries depend on the solid cylinder diameters (d) and heights (h). The solid cylinder diameters used are 1.5 mm and 3 mm and the heights 0.75 mm and 1.5 mm respectively. The 3 mm diameter and 1.5 mm height solid cylinder has the same dimensions as the magnet used to perform the experimental measurements. This last volume value was simulated to compare the experimental results.

**RESULTS AND DISCUSSION**

First the vibration of a bare clamped membrane under an in-plane uniform force per unit area, exerted on its four sides as shown in Figure 4, was simulated to be sure that we obtain the in-plane force frequency dependence on the stress caused by a uniform in-plane force, which for a rectangular membrane is described by (1). The dependence of the resonance frequency on the stress in this case is shown in Figure 5 (a). This curve fits, as expected, to a square root function. Besides, as it can be seen from Figure 5 (b) when focusing on low force values from zero to 45 mN, the dependence is linear and the obtained curve fits to $(f - f_0) = 7524 Hz N^{-1} F$ with a coefficient of determination $R^2 = 0.9998$, where $F$ is the applied force and $f_o$ is the first mode resonance frequency without applied force. This linear region is the one used by sensors based on vibrating membranes to carry out the measurements. Therefore, these results confirm that the simulation reproduces the behaviour of a rectangular vibrating membrane under an in-plane uniform force.

The effect of an out-of-plane force on a four side clamped membrane was then analysed using both, a punctual force and a surface force applied on a certain surface of the membrane. The former case was simulated using a punctual out-of-plane force, applied right on the centre of the membrane, while for the latter case the simulation was carried out using a solid cylinder to determine the surface where the out-of-plane force was applied (see figure 3) .

The resulting curve shown in Figure 6 shows the results obtained for the punctual force simulation. The curve can be split into two different regions. It starts with a quadratic dependence of the frequency from zero to 2 mN, turning into a square root dependence for higher force values. Figure 7 (a) shows the quadratic behaviour, fitting the curve to a polynomial of the shape $(f - f_0) = 1.2 \cdot 10^8 \, Hz N^{-2} F^2$, with $R^2 = 0.999$, while Figure 7 (b) shows the square root dependence which fits, with $R^2 = 0.9997$, to $(f - f_0) = -3000 Hz + 7.13 \cdot 10^4 \, Hz N^{-1/2} \sqrt{F}$ . It means that the membrane, under an out-of-plane force, behaves as a membrane stressed by means of an in-plane force, from a certain force value.

The fact that the membrane has a solid cylinder on its surface makes all the points in contact with the solid surface vibrate in a coplanar way. Since the constant force will be applied only on the solid surface, it means that when the solid surface increases, the vibrating free-surface of the membrane decreases. Thus, the different geometries of the membrane are related with the vibrating free-surface of the membrane. Since this fact should affect the behaviour of the resonance frequency, a vibrating membrane with solid cylinders of different volumes was simulated. First a membrane with a solid cylinder of 3 mm diameter and 1.5 mm height, fixed and centred on its surface was simulated under an out-of-plane force. The solid cylinder dimensions were chosen in order to compare the results with some experimental measurements of the frequency dependence on the out-of-plane force reported elsewhere [4]. The curve obtained from the simulations shows the same behaviour as the one shown in Figure 7 from the punctual out-of-plane force effect, but in this case the quadratic dependence has a wider force range [0-8] mN, turning from this last force value to a square root dependence. To complete the study, some simulations were performed with a solid cylinder of 1.5 mm diameter and 0.75 mm height. Figure 8 shows the three simulated curves obtained and table 3 shows the coefficient values of the quadratic fits ($\alpha$), and the quadratic range in every case. Figure 8 (a) shows how the rate of change of the frequency depends on the solid cylinder diameter, and therefore on the membrane vibrating free-surface. It shows up the fact that the larger the diameter, the lower the frequency rate of change, that is, the smaller the quadratic coefficient as is shown in table 3. However, from Figure 8 (b) it can be seen how, as the diameter increases, the range where the quadratic dependence of the frequency is found, increases. From the Table 3 it can be seen that the punctual force and the solid cylinder with d=1.5 mm present similar quadratic ranges, increasing slightly from one to another. When the diameter is close to the membrane frame, the free vibrating surface decreases considerably (1 mm is the minimum distance from the frame of the membrane to the edge of the 3mm diameter solid) and the increment of the quadratic range is much more marked. A higher magnet diameter means more surface of the membrane under the effect of the applied force and therefore, less free vibrating surface. The larger the surface of the membrane in contact with the magnet, the higher the quadratic range, but the lower the variation of the frequency in the same force range.

The deformation of the membrane as a function of the applied force was simulated for the three different geometries. All the cases report a linear dependence of the deformation from zero to a certain force value, which changes for every geometry. From this force value the non-linear geometric phenomenon start to appear for the membrane deformation. The linear range for the deformation fits exactly with the quadratic range for the frequency dependence. That is, when the relation deformation-force is linear, the frequency dependence shows a quadratic response as a function on the force, and starts to behave as a membrane stressed by means of an in-plane force when the non-linear geometric phenomenon appears.

Therefore, the linear range of the deformation increases when increasing the surface of the solid. The more the coplanar vibrating-points of the membrane, the higher the linear range of the deformation and the higher as well the quadratic range of the frequency dependence.

To complete this study the simulated results were compared with the experimental results obtained using a gradiometer based on a vibrating membrane with a fixed magnet centred on its surface [4]. In this device, as it was explained above in the introduction section, an out-of-plane force is applied, giving raise to a change in the resonance frequency. The graph shown in Figure 9 is plotted in force terms (magnetic force), taking into account the relation between the value of the magnetic force and the value of the magnetic field gradient :

$$F_z = m_z \left( \frac{\partial B_z}{\partial z} \right) \qquad (2)$$

where $F_z$ is the out-of-plane force, $m_z$ the out-of-plane component of the magnetic moment of the magnet (which is known to be oriented along the out-of-plane direction and whose value is 6.75 10$^{-3}$ Am$^2$) and *($\partial B_z/\partial z$)* the *$B_z$*-gradient along the out-of-plane direction. In order to compare the dependence on the out-of-plane force applied on the solid cylinder or the magnet surface, both experimental and simulated curves have been plotted in the same graph. The simulated curve, as it has been explained before is the result of simulating the vibration of the Si (100) membrane, using the clamped boundary conditions, with a fixed SmCo magnet centred on its surface, under an applied out-of-plane force (negative in the gravity direction). The simulation does not take into account the magnet weight in order to reduce the computation time. Therefore the simulation results shown in Figure 9 are shifted introducing this value to make easier the comparison. Not the complete device but only the membrane with the solid cylinder was simulated. The frequency resonance is shown as the difference between the measured or simulated frequency for every force value and the minimum value of frequency obtained respectively. It can be seen the good agreement between them, confirming the validity of the experimental results. The experimental curve shows a pure quadratic behaviour in the force range of [0-7] mN. Taking into account the fact that the simulations report a quadratic behaviour in the force range of [0-8] mN, the pure quadratic behaviour measured experimentally now can be understood.

Focusing on the gradiometer response, if the frequency dependence on the total applied force is quadratic:

$$(f - f_0) = \alpha \cdot F^2 \tag{3}$$

then, the sensitivity can be expressed:

$$\frac{\Delta(f - f_0)}{\Delta F} = 2\alpha \cdot \left(F_g + F_{ext}\right) \tag{4}$$

being *$\alpha$* the quadratic coefficient shown in table 3, and taking into account that the total force is the sum of the weight or the gravity force ($F_g$) and the external force applied on the system ($F_{ext}$). From these results it is clear that the weight of the magnet improves the sensitivity of the system shifting the curve to higher slopes.

On the other hand, to understand the effect of the volume of the magnet used on the gradiometer sensitivity, it is interesting to plot the calculated frequency dependence for the two solid cylinder used, as a function of the magnetic field gradient, considering the magnetic moment of every solid cylinder depending on its volume. If we assume the solid cylinders are SmCo magnets with a magnetization value of  M = 0.8 T, the magnetic moment is easily calculated knowing the volume of the magnet. The calculated frequency dependence on the magnetic field gradient for the two different solid cylinders as SmCo magnets is shown in Figure 10.

If we focus in the results of Figure 8 where the frequency dependence is expressed as a function of the applied out-of plane force, being the force the same for the two solid cylinders, it can be seen that a magnet with the less possible diameter value would be desirable, since a reduction of this parameter means a larger slope in the resulting curve and therefore, more sensitivity. Since the magnetic field gradient depends on the magnetic moment of the magnet used for the sensor head of the gradiometer and as it can be seen from Figure 10, the magnet with the larger volume gives raise to a higher sensitivity in spite of the decreasing of the membrane free-vibrating surface.

As it has been explained, the frequency dependence shows two regions where the measurements can be carried out. The first region (for low forces), is the quadratic one which presents a higher sensitivity detection zone. Nevertheless, there is another region which shows a square root dependence and therefore is a lower sensitivity detection zone, but it is a region which allows the measurements for higher force ranges.

Therefore, the magnet used for the sensor head of the gradiometer should be as large and heaviest as possible. This fact will be taken into account in future experiments. Taking into account the punctual force results, a magnet with the least diameter value but with the largest magnetic moment, that is, with the largest height without creating additional stresses due to a tilt of the magnet while vibrating, is the best option. Some micromagnets are currently being developed for this purpose, since the commercial magnets do not present the desirable dimensions.

**CONCLUSIONS.**

Since a lot of devices base its detection in the vibration of a silicon membrane, the understanding of the effect of the out-of-plane forces is important to improve and optimize the response signal of these devices. In this work it has been proved that when a membrane is stressed by means of an out-of-plane force, the dependence of the frequency becomes quadratic from zero to a certain value of the force turning to a square root dependence as in the in-plane force case. This value of the force, from which the dependence changes, depend on the free vibrating surface of the membrane and is higher when the free vibrating surface decreases. The rate of change of the resonance frequency depends as well on the free vibrating surface, being higher when it increases. The linear range of the membrane deformation is directly related with the vibrating free-surface of the membrane, fitting exactly with the quadratic range of the frequency. The more the coplanar vibrating points the membrane contains, the higher both ranges are. The experimental results showing a quadratic behaviour have been demonstrated to be consistent with the theory by means of simulations.

**ACKNOWLEDGEMENTS**

This work has been partially financed by the project OPTOMAG-MANTIS (ESP2005–05278 of the Spanish Space National Program, granted by the Ministerio de Ciencia e Innovación.

**TABLE CAPTIONS**

Table 1. Mechanical properties of silicon membranes and the solid cylinders used in the simulations.

Table 2. Simulated membrane geometries.

Table 3. Coefficient values of the quadratic fits and quadratic range for the three different solid cylinder diameters.

**FIGURE CAPTIONS**

Figure 1. Photographic detail and (inset) schematic view of the experimental set-up.

Figure 2. Scheme of the vibration of the membrane with the magnet fixed on it. (a) Without vibrating. (b) Vibrating because of the ac magnetic field created by the excitation coils. (c) Effect of an external magnetic force.

Figure 3. First mode of vibration of the square simulated membranes. Bare (left) and with a fixed solid cylinder (right), centred on its surface.

Figure 4. Square membrane put down a uniform in-plane force.

Figure 5. Simulation (dots) and mathematical fit (line) of the membrane put down an in-plane force. (a) square root dependence and (b) linear dependence. $f_o = 4741$ Hz.

Figure 6. Simulation of an out-of-plane punctual force applied on the centre of the membrane. $f_o = 8693$ Hz.

Figure 7. Simulation of an out-of-plane punctual force applied on the centre of the membrane. Simulation (dots) and mathematical fit (line). (a) quadratic dependence and (b) square root dependence.

Figure 8. Comparison of the simulations of a punctual force (black circles. $f_o = 8693$ Hz.), force applied on a solid cylinder of 1.5 mm diameter and 0.75 mm height (white circles. $f_o = 1637$ Hz.) and 3 mm diameter and 1.5 mm height (triangles. $f_o = 1313$ Hz.). (a) Comparison in the range [0-25] mN and (b) Comparison in the range [0-16] mN.

Figure 9. Experimental (triangles) and simulated results (circles) of the membrane with a magnet of 3 mm diameter centred on its surface. $f_{min}$ is the minimum frequency in every case, measured and simulated. $F_{ext}$ is the applied magnetic force. The sensor head of the device (a membrane with a magnet fixed on its surface) is shown in the graph (down right). The sensor head is fixed to a holder.

Figure 10. Simulated frequency dependence on an out-of -plane magnetic field gradient applied on a SmCo magnet of 1.5 mm diameter and 0.75 mm height (white circles) and 3 mm diameter and 1.5 mm height (triangles).

|  | $E_x$ (Pa) | $E_y$ (Pa) | $E_z$ (Pa) | $\nu_{xy}$ | $\nu_{xz}$ | $\nu_{yz}$ | $\rho$ (Kg/m$^3$) |
|---|---|---|---|---|---|---|---|
| Silicon | $1.69\ 10^{11}$ | $1.69\ 10^{11}$ | $1.30\ 10^{11}$ | 0.0623 | 0.2784 | 0.2784 | $2.33\ 10^3$ |
| Solid cylinder | $1.50\ 10^{11}$ | $1.50\ 10^{11}$ | $1.50\ 10^{11}$ | 0.3 | 0.3 | 0.3 | $7.61\ 10^3$ |

Table 1.

| Punctual force | Solid cylinder d = 1.5 mm h = 0.75 mm | Solid cylinder d = 3 mm h = 1.5mm |
|---|---|---|
| 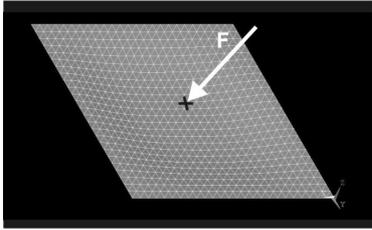 | 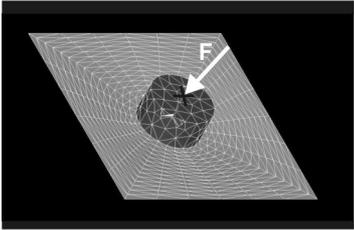 | 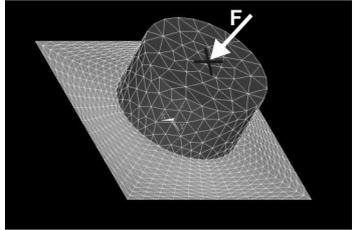 |

Table 2.

|  | Punctual force | Solid cylinder d = 1.5 mm | Solid cylinder d = 3 mm |
|---|---|---|---|
| Quadratic range (mN) | [0,2] | [0,3] | [0,8] |
| $\alpha$ (Hz/N$^2$) | $1.2\ 10^8$ | $1.6\ 10^6$ | $6.1\ 10^5$ |

Table 3.

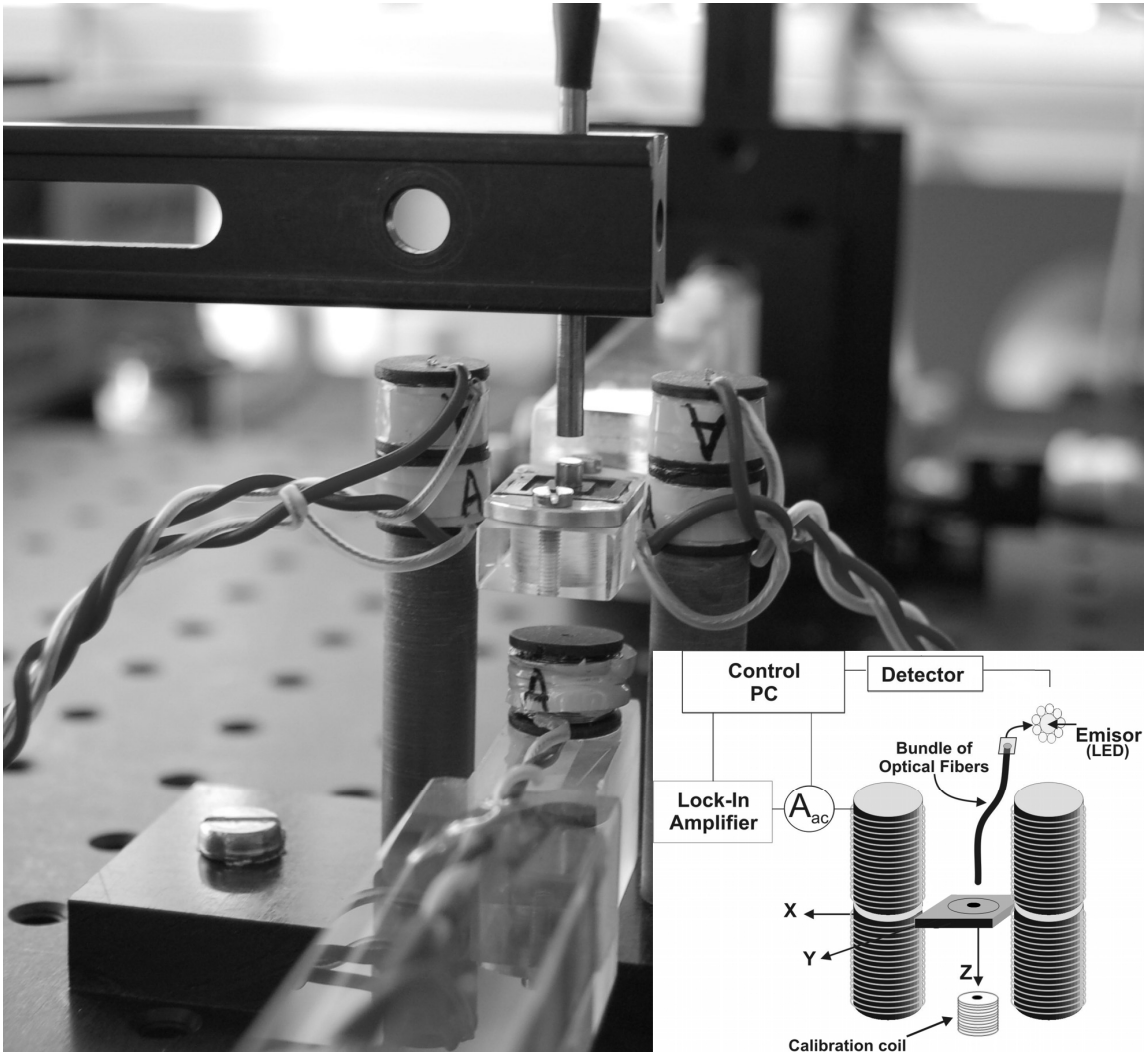

Figure 1.

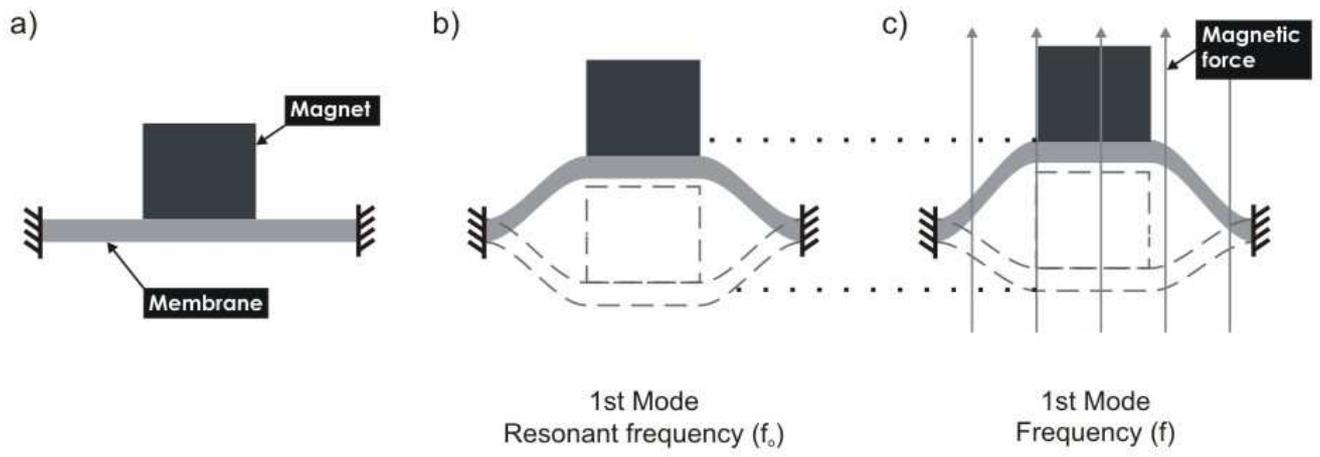

a)

Magnet

Membrane

b)

1st Mode
Resonant frequency ($f_o$)

c)

Magnetic force

1st Mode
Frequency (f)

Figure 2.

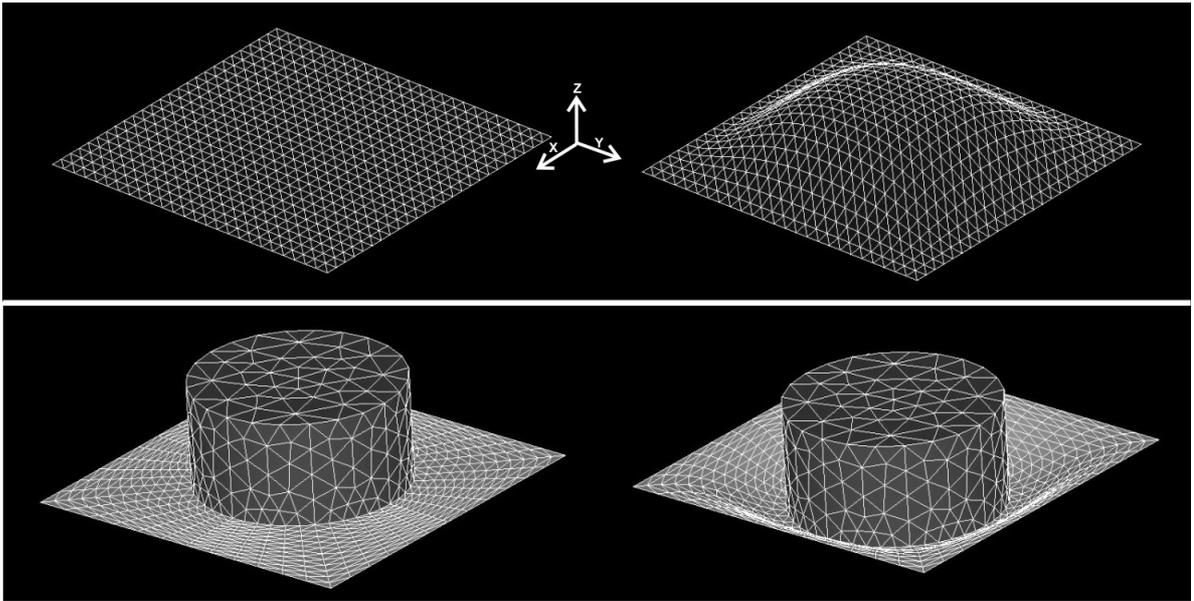

Figure 3.

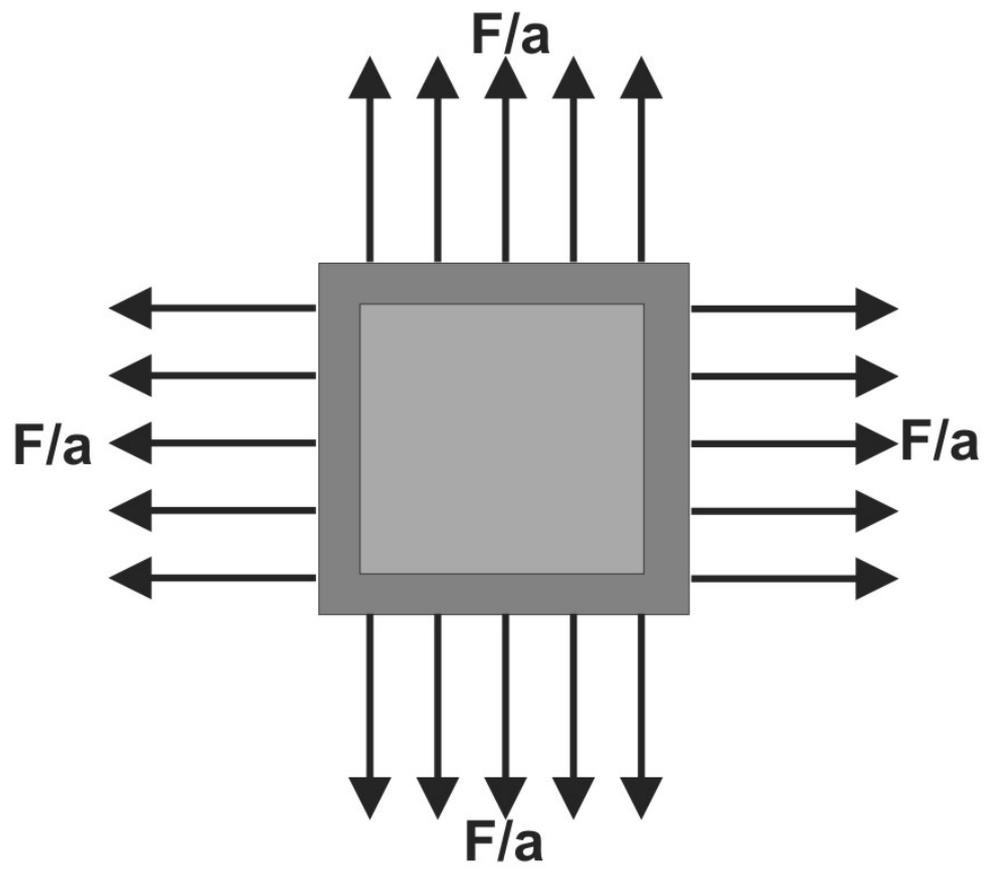

Figure 4.

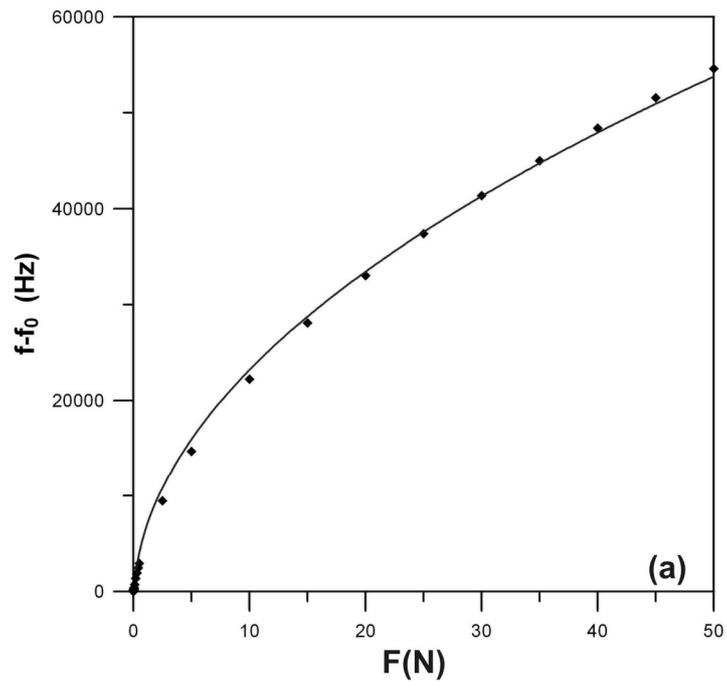

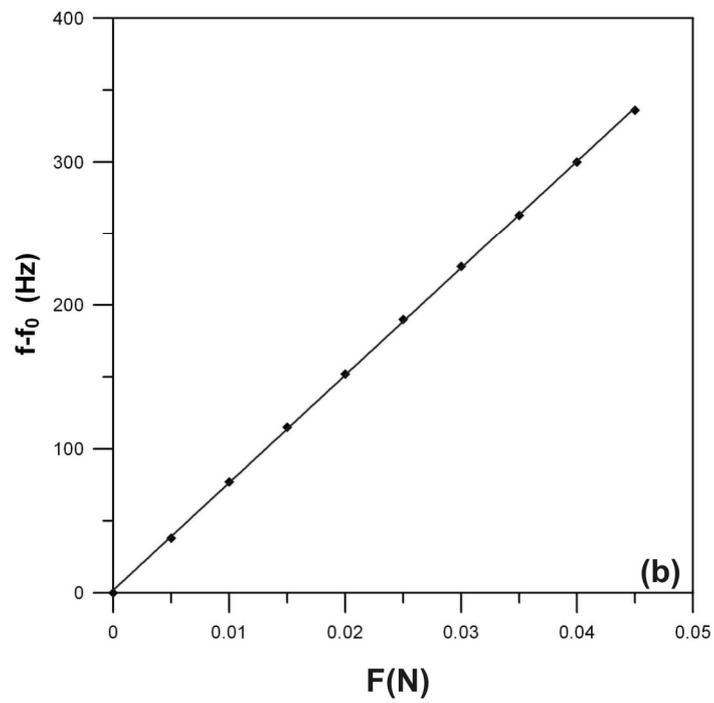

Figure 5.

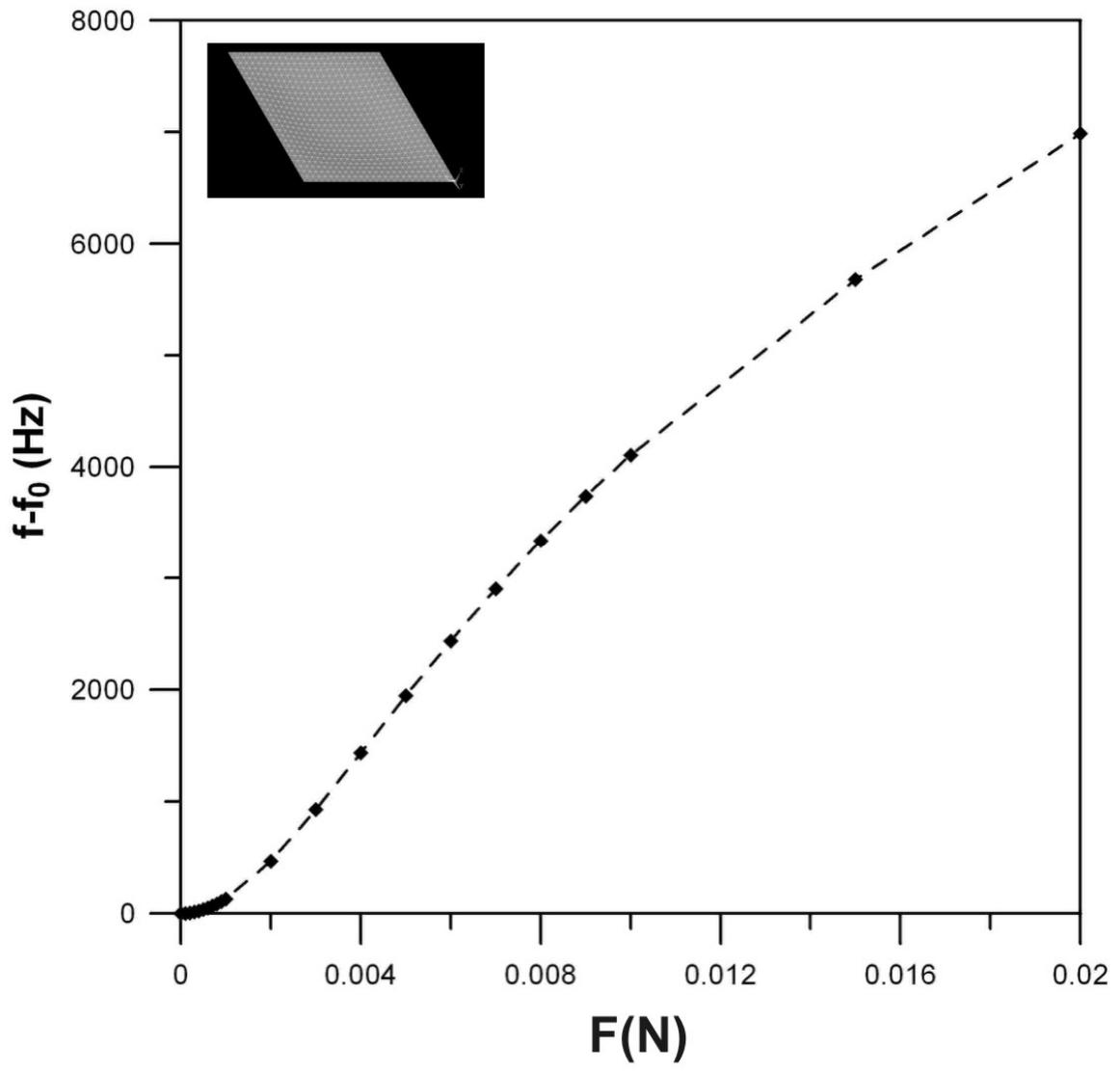

Figure 6.

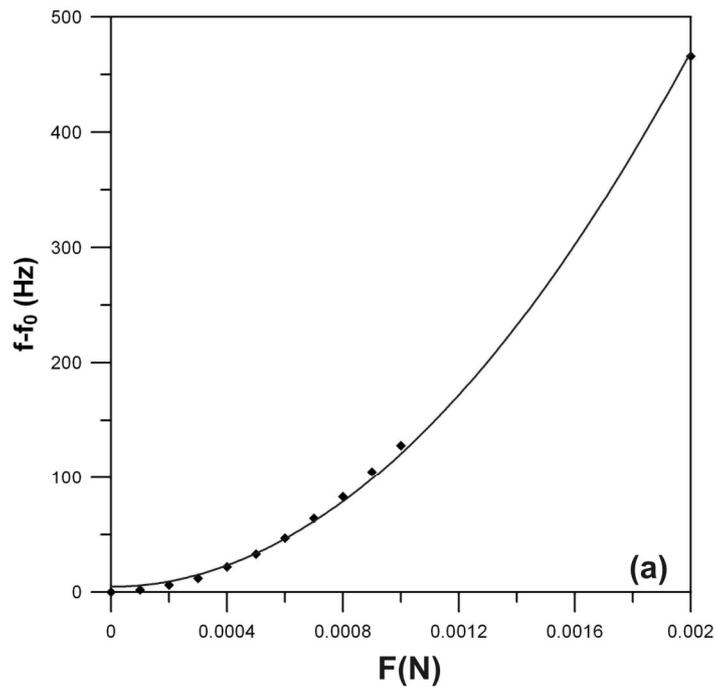

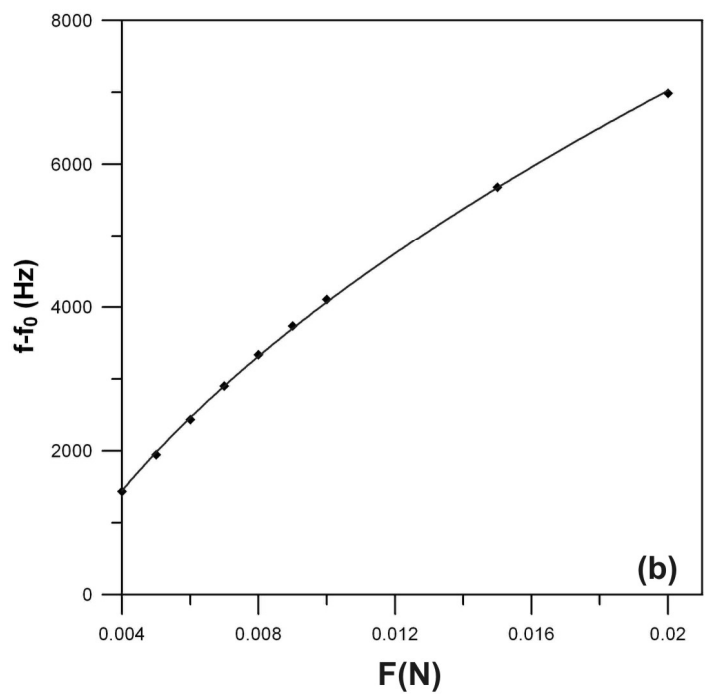

Figure 7.

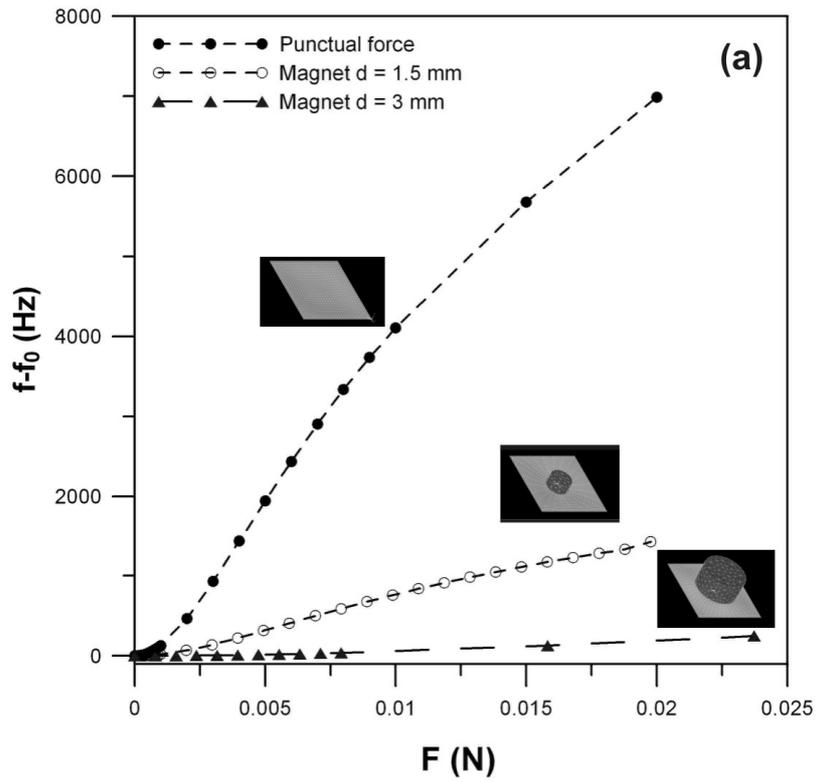

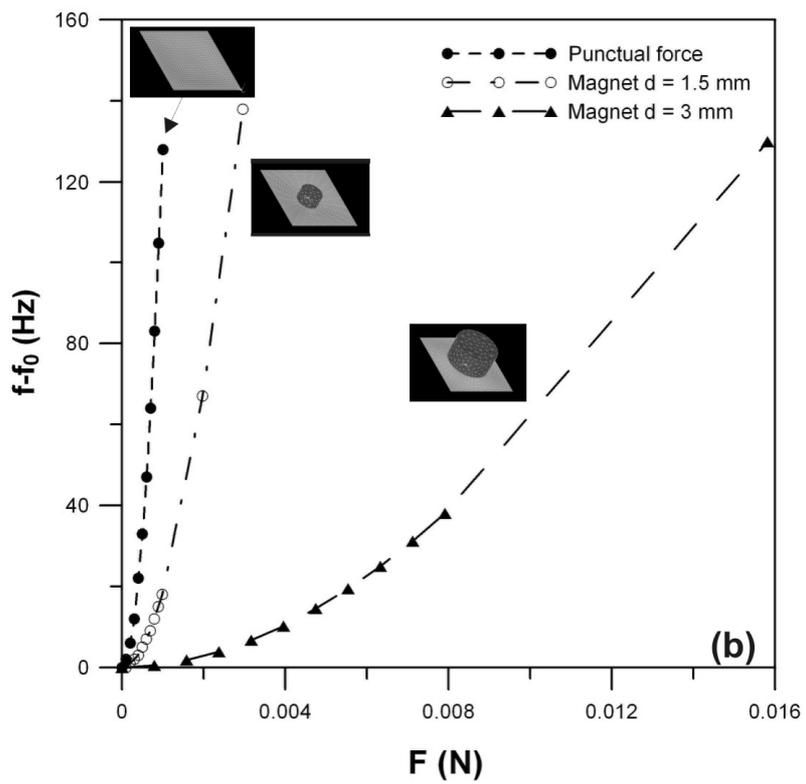

Figure 8.

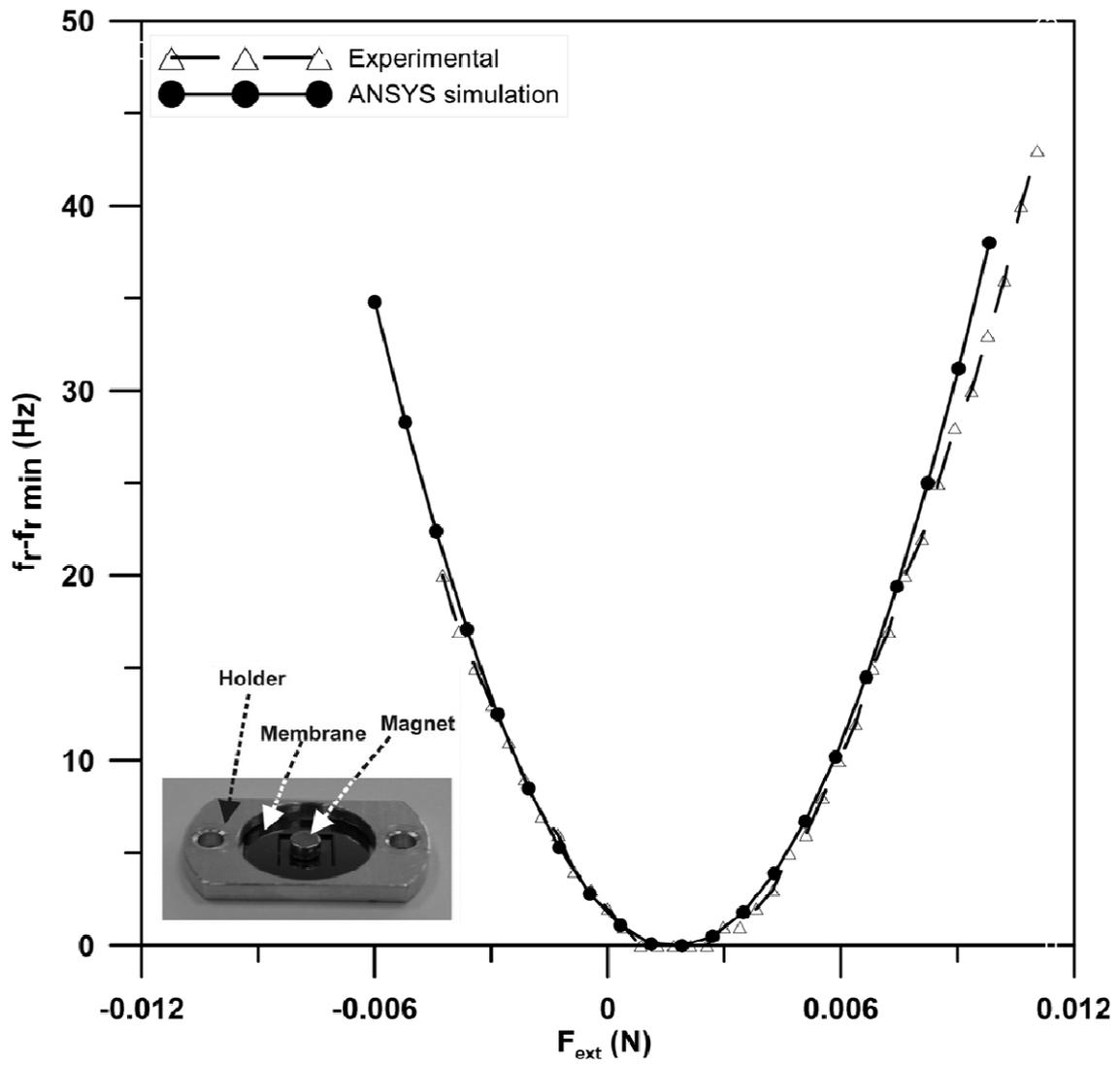

Figure 9.

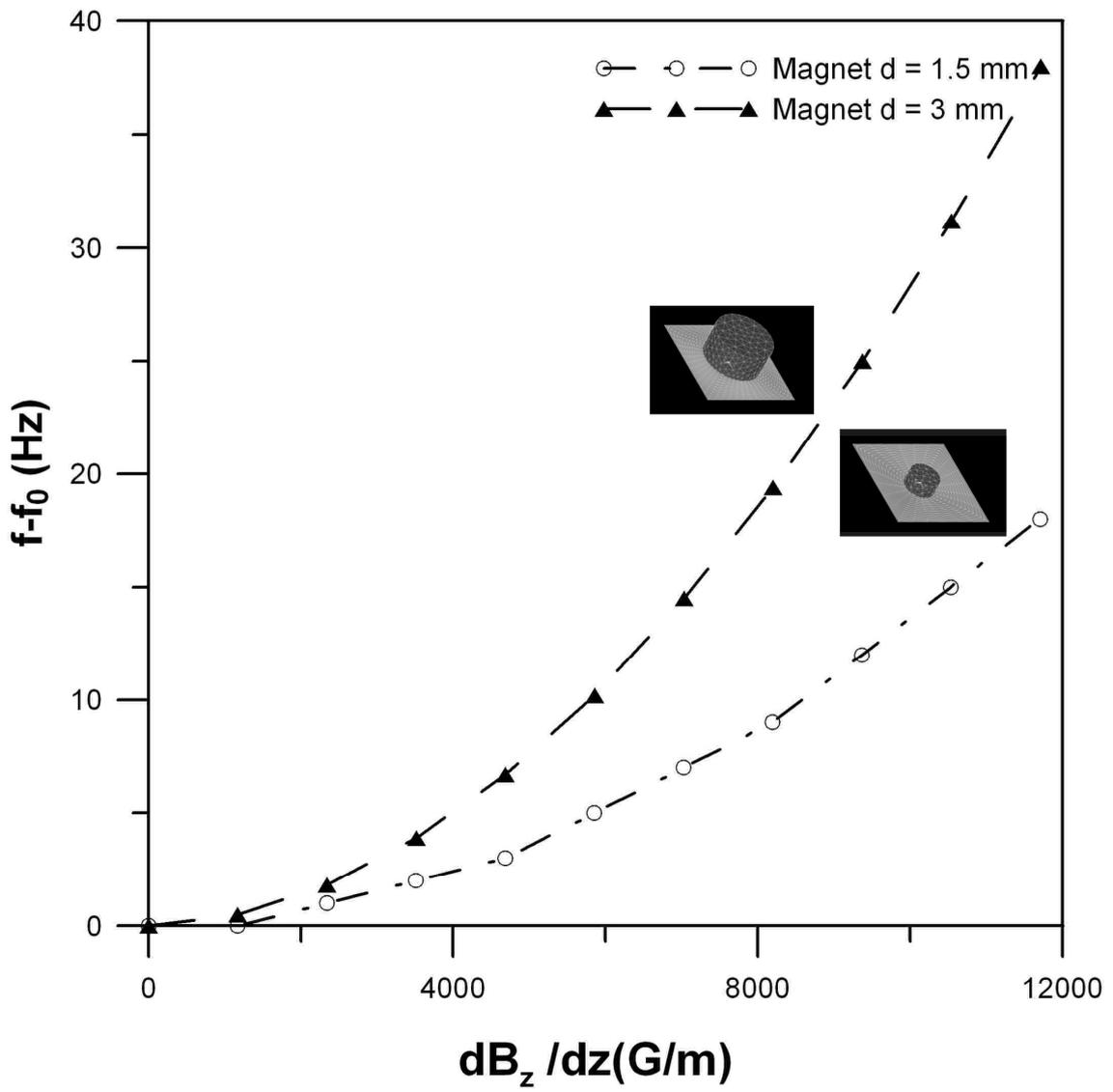

Figure 10.

**Irene Lucas** (Ciudad Real, Spain, 1978), MSc (2003) in Physical Sciences and PhD (2009) by the Universidad Complutense de Madrid (UCM) and PhD (2009) by the Universidad Complutense de Madrid (UCM). During the PhD, her research was focused on growth and characterization of soft and hard magnetic materials, and magnetic sensors. She is currently working as a Marie Curie postdoctoral researcher in the Leibniz Institute for Solid State and Materials Research (IFW) in Dresden, Germany. Her research is currently focused in high temperature superconductors.

**Rafael Pérez del Real** (Madrid, Spain, 1966), MSc (1989) in Physical Sciences and PhD (1993) by the Universidad Complutense de Madrid (UCM). His research has been focused on soft magnetic materials, magneto-optics, magnetic materials for cancer treatment, and magnetic sensors. He is currently working in the Materials Science Institute of Madrid (CSIC) on magnetic nanomaterials.

**Marina Díaz-Michelena** (Madrid, Spain, 1975) MSc (1998) in Physical Sciences by the Universidad Complutense de Madrid (UCM) and PhD (2004) in Applied Physics in the Universidad Poletécnica de Madrid (UPM). She is researcher in magnetic sensors for Space applications at INTA, the Spanish National Institute of Aerospace Technology. Associate Lecturer in the Department of Physics of the Materials of the Faculty of Physics of UCM.

**V. De Manuel** (Madrid, Spain, 1981), MSc (2004) in Physical Sciences by the Universidad Autónoma de Madrid (UAM) and PhD (2009) by the Universidad Complutense de Madrid (UCM). His PhD work was focused in soft magnetic amorphous and nanocrystalline materials and magnetic sensors

**Marta Duch** was born in Barcelona, Spain in 1967. She received her Technical Engineer degree in Chemistry in the Universitat Politècnica de Catalunya, Barcelona, Spain (1992). She is currently working in the Micro- and Nanosystems Department at the Instituto de Microelectrónica de Barcelona IMB-CNM (CSIC). Her main area of activity is focused on bulk and surface silicon micromachining, glass etching and chemical and electrochemical deposition of metals.

**Jaume Esteve** was born in Parets del Vallés, Barcelona, Spain, in 1961. He received the BSc and thePhDdegrees in physical electronics from the Universitat de Barcelona in 1984 and 1988, respectively. In 1990, he joined the Department of Silicon Technology and Microsystems, at the Instituto de Microelectrónica de Barcelona IMB-CNM (CSIC), as a senior research scientist. His areas of interest include silicon micromachining technologies and their application to integrated sensors and actuators. Pr. Esteve holds six patents and has published more than 70 research papers, and he has served as a reviewer of several international scientific journals, and as a referee for public-funded projects.

**José A. Plaza** was born in Cerdanyola del Vallés (Barcelona), Spain in 1968. He received his physicist degree and his PhD degree in Electronics engineering from the Universitat Autònoma de Barcelona, Spain (1992, 1997). He has the degree of Specialist in simulation by the Finite-Element Method from the Universidad Nacional de Educación a Distancia since 1995. He is currently working in the Micro and Nanosystems Department at the Instituto de Microelectrónica de Barcelona IMB-CNM (CSIC). He has been focused on technology development, design, simulation and characterization of microsystems and nanosystems. Now, his research is focused on the fabrication of MEMS/NEMS to study living cells. Dr. Plaza has served as a reviewer of several international scientific journals, and as a referee for public-funded projects in Spain.